\documentclass[twocolumn,aps,showpacs,prl]{revtex4}
\usepackage[dvips]{graphicx}

\begin{document}

\title{Flexoelxctricity of the distorted twist bend nematic phase}

\author{I. Lelidis and E. Kume}
\affiliation{
Faculty of Physics, National and Kapodistrian University of Athens,\\
Panepistimiopolis, 15784 Zografos, Athens, Greece
}
\date{\today}

\begin{abstract}
Mesogenic dimers in the twist-bend nematic phase exhibit much higher flexoelectric polarization than in their uniform nematic phase. In order to theoretically investigate this data, we extend the symmetry based linear elastic theory of the twist-bend nematic phase developed in Phys. Rev. E 92, 030501(R) 2015, by including flexoelectricity under the action of an external electric field perpendicular to the helical axis. We show that at the nematic towards twist-bend nematic transition, a new flexoelectric mode becomes active. Consequently, the present model predicts the increase of the effective flexoelectric coefficient when the system is entering the twist-bend nematic phase. The influence of the flexoelectric coupling on the equilibrium wavevector  and the spontaneous heliconical tilt angle are investigated. The electroclinic coefficient is calculated. Finally we argue that the helix could be unwound giving rise to a splay-bend nematic phase.

\end{abstract}

\maketitle

%%%%%%%%%%%%%%%%%%%%%%%%%%%%%%%%%%%%%%%%%%%%%%%%%%%%%
\section{Introduction}
	
Flexoelectricity was first discussed for crystals by Kogan \cite{kogan}. In nematic liquid crystals, flexoelectricity describes the coupling between the nematic director distortions and electric polarization, and it was first discovered by Meyer \cite{mayer68,mayer69a,mayer69b,luiz,buka}. In contrast to piezoelectricity, flexoelectricity does not require the lack of inversion symmetry and therefore the effect is observed both in chiral and in achiral liquid crystals (LC).

Nematic LCs exhibit long range orientational order which is characterized by the nematic director ${\bf n}$, coinciding with the statistical average of the molecular long axis for rod-like molecules \cite{prost}. The nematic phase is apolar, in the sense that ${\bf n}$ is equivalent to $-{\bf n}$, and hence the electric polarization, in the non-deformed state is identically zero. However when the nematic is distorted, it appears an electric polarization ${\bf P}$ proportional to the local deformation of the director field, $n_{i,j}=\partial n_i/\partial x_j$, that is, $P_i=\lambda_{ijk} n_{j,k}$, where $\lambda_{ijk}$ are the elements of the flexoelectric tensor, $n_i$ the cartesian components of ${\bf n}$, and $x_j$ the cartesian components of the positional vector ${\bf x}$ \cite{mayer69b}.

For cholesteric liquid crystals, the application of an electric field perpendicular to the helix axis leads to the rotation of the optical axis arising from flexoelectric coupling which is linear in the electric field \cite{patel}. Flexoelectricity has been also discovered  in the twist-bend nematic ($N_{TB}$) phase \cite{claire,balachandran}. Recent studies show that the flexoelectricity in the uniform nematic phase of dimeric molecules is stronger than in monomeric nematics while in the $N_{TB}$-phase of the same compound flexoelectric polarization doubles its amplitude \cite{ferrarini,balachandran,Zhou}.
	
Recently we proposed an elastic continuum theory for the $N_{TB}$-phase \cite{pre} where it was shown that the existence of a twist-bend deformation in the ground state can be justified assuming that the elements of symmetry of the phase are the nematic director ${\bf n}$ and the helical axis unitary vector ${\bf t}$. Assuming that the twist-bend nematic phase is not spondaneously ferroelectric, the flexoelectric tensor was calculated in \cite{lc2016}. In the ground state, the new terms of the flexoelectric tensor, in respect to the nematic phase, are identically zero \cite{lc2016}. Here we extend this model to describe flexoelectric polarization of a twist-bend nematic phase under the action of an electric field perpendicular to the helix axis. We show that at the $N-N_{TB}$ transition a new flexoelectric contribution appears caused by the electric field induced distortion of the heliconical tilt angle. We calculate the rotation angle between the optical axis of the structure and the helicoidal axis as function of the electric field amplitude and the effect of this rotation on the twist-bend cone angle as well as to the wavevector of the structure. Finally, we discuss some experimental data in respect to our model.

\section{Model}

The elastic free energy of a twist-bend nematic is characterized by two unitary directors, the nematic director ${\bf n}$ and the helix axis ${\bf t}$ related to the collective twisted  arrangement of the molecules. As shown previously \cite{pre}, the free energy density compatible with the symmetry of the phase, in the absence of an external field, may be written as

\begin{widetext}
	\begin{eqnarray}
		\label{fzao}
		f_{e\ell} &=& f_0 - \frac{1}{2} \eta ( {\bf n} \cdot {\bf t})^2  + \kappa_1 \,{\bf t}\cdot \left[{\bf n}\times(\nabla \times {\bf n})\right]+\kappa_2\,{\bf n}\cdot (\nabla \times {\bf n}) \nonumber + \kappa_3 ({\bf n}\cdot{\bf t}) (\nabla \cdot{\bf n})
		+ \frac{1}{2}K_{11} (\nabla \cdot {\bf n})^2 \nonumber \\
		&+& \frac{1}{2} K_{22} \left[{\bf n} \cdot (\nabla \times {\bf n})\right]^2
		+ \frac{1}{2} K_{33} ({\bf n} \times \nabla \times {\bf n})^2
		- (K_{22} + K_{24}) \nabla \cdot ({\bf n} \nabla \cdot {\bf n} + {\bf n} \times \nabla \times {\bf n}) \nonumber \\
		&+& \mu_1 [{\bf t} \cdot ({\bf n} \times \nabla \times {\bf n})]^2 + \nu_1 [{\bf t} \cdot \nabla ({\bf t} \cdot {\bf n})]^2
		+ \nu_2 [{\bf t} \cdot \nabla ({\bf n} \cdot {\bf t}) (\nabla \cdot {\bf n})]
		+ \nu_3 [\nabla ({\bf t}\cdot{\bf n})]^2 + \nu_4 [({\bf t} \cdot \nabla) {\bf n}]^2 \nonumber \\
		&+& \nu_5 [\nabla ({\bf n} \cdot {\bf t})\cdot ({\bf t} \cdot \nabla) {\bf n}] + \nu_6 \nabla ({\bf n}\cdot{\bf t})\cdot (\nabla \times {\bf n}). \nonumber \\
	\end{eqnarray}
\end{widetext}
where $\kappa_i$, for $i=1,2,3$, are the elastic constants connected with the spontaneous splay, twist and bend, respectively, $K_{11}$, $K_{22}$, $K_{33}$ and $K_{24}$ are the Frank elastic constants \cite{prost}. $\eta$ is a measure of the coupling strength between the two directors $\bf n$ and $\bf t$. $\mu_1$ and $\nu_j$, for $j=1,2,..., 6$ are elastic constants.

In the $\rm N_{\rm TB}$ phase, the  nematic director $\mathbf{n}$ forms an oblique helicoid around the direction
$\mathbf{t}$ of the helix axis, with a constant tilt angle $\theta$. By taking ${\bf t} = {\bf e_z}$, the nematic director is written as

\begin{equation}
 	\label{tbkateuthintis}
 	\mathbf{n}=[\cos\phi(z)\mathbf{e_x}+\sin\phi(z)\mathbf{e_y}]\sin\theta+\cos\theta\mathbf{e_z}
 \end{equation}
where $\phi (z)$ is the azimuthal angle.

The flexoelectric polarization of the $N_{TB}$ phase generally includes two contributions. The first one, is the usual flexoelectric polarization of conventional nematics and cholesterics

	\begin{equation}
		\mathbf{P_n}=e_1\mathbf{n}(\nabla\cdot\mathbf{n})-e_3\mathbf{n}\times(\nabla\times\mathbf{n})
	\end{equation} where $e_{1}$ and $e_{3}$ are the flexoelectric coefficients for splay and bend respectively \cite{mayer68,prost}. In the $N_{TB}$ ground state  the splay term is identically zero, that is, the spontaneous flexoelectric polarization has only a bend type contribution. Nevertheless, under an electric field the $N_{TB}$ helix is distorted and therefore a non zero splay appears.
As discussed in \cite{lc2016}, a second contribution $\mathbf{P_{t}}$  to the flexoelectric polarization arises from the coupling of the helix axis $\mathbf{t}$ with the nematic director $\mathbf{n}$
\begin{equation}\label{totalfl}
		\mathbf{P_{t}}=-\beta_1\mathbf{t}\{\mathbf{t}\cdot[\mathbf{n}\times (\nabla\times\mathbf{n})]\}+\beta_3\mathbf{n}[\mathbf{t}\cdot\nabla(\mathbf{n}\cdot\mathbf{t})]
	\end{equation}
$\mathbf{P_{t}}$ is identically zero in the absence of the $N_{TB}$ cone distortions. Therefore, the total flexoelectric polarization
$\mathbf{P_{fl}}=\mathbf{P_n}+\mathbf{P_t}$, in the absence of distortions, reduces to
\begin{equation}
		\mathbf{P_{fl}}(E=0)=\frac{qe_3}{2}\sin(2\theta)[-\sin\phi(z)\mathbf{u_x}+\cos\phi(z)\mathbf{u_y}]
\end{equation} and it remains perpendicular both to the nematic director and to the helix axis while it rotates along the helix. Nevertheless, the total flexoelectric polarization for a period along the helix integrates to zero. Therefore, no macroscopic flexoelectric polarization survives in an undistorted  $N_{TB}$ phase. Finally, the total energy density of the system, including dielectric coupling, is written as

\begin{equation}
f_{tb}=	f_{e\ell}+f_{f\ell}+f_{diel}=f_{e\ell}-\mathbf{E}\cdot\mathbf{P_{fl}}-\frac{1}{2} \varepsilon_0\varepsilon_a\,(\mathbf{E\cdot n})^2
\end{equation}
where $\varepsilon_0$ is the permittivity of the free space and $\varepsilon_a$ is the local dielectric anisotropy of the liquid crystal compound. For high enough electric field the dielectric term dominates over the flexoelectric one.

In the following, to favor the flexoelectric torque, we consider a distorted $N_{TB}$ phase where the distortion is induced by an electric field in the $x-$direction perpendicular to the helix axis, $\mathbf{E}=E\mathbf{e_x}\perp \mathbf{t}$. As it is well known, the flexoelectric torque provokes a rotation of the optical axis $\mathbf{N}$ by an angle $\alpha$ in the $(y,z)$ plane and results to the following expression for the nematic director
	\begin{equation}
		\mathbf{n}(z)=\left( \begin{array}{c}
	\cos\phi\sin\theta	\\
		\cos\alpha\sin\phi\sin\theta-\sin\alpha\cos\theta 	\\
		\sin\alpha\sin\phi\sin\theta+\cos\alpha\cos\theta	\end{array}\right)
	\end{equation}
	
Using the latter expression for the stressed director, we calculate after some algebra the mean energy density per pitch $p$ defined by
	\begin{equation}
		<f_{tb}>=\frac{1}{p}\int^p_0{f_{tb}\mathrm{d}z}
	\end{equation}Under the usual assumption that $\phi=qz$ with $q=2\pi/p$ the  wavevector  and $p$ the pitch of the modulation,  $<f_{tb}>$ is given from the following expression	
{\small \begin{eqnarray}\label{fmean}
 <f_{tb}> &=& f_0-\frac{1}{2}\eta\left( \frac{1}{2}\sin^2\alpha\sin^2\theta+\cos^2\alpha\cos^2\theta\right) \nonumber\\
&-& \kappa_2q\cos\alpha\sin^2\theta+\nu_4 q^2\sin^2\theta\nonumber\\
&+&\frac{1}{4}\widetilde{K}_{11}q^2\sin^2\alpha\sin^2\theta \nonumber\\
&+&\frac{1}{2} K_{22}q^2\,\sin^2\theta \left(\cos^2\alpha\sin^2\theta +\frac{1}{2}\sin^2\alpha\cos^2\theta\right) \nonumber\\
&+&\frac{1}{2}K_{33} q^2\,\sin^2\theta\left( \cos^2\alpha\cos^2\theta +\frac{1}{2}\sin^2\alpha\sin^2\theta\right)\nonumber\\
&+&\frac{1}{2}\mu_1 q^2\,\sin^2\alpha\sin^2\theta\left( \cos^2\alpha\cos^2\theta +\frac{1}{4}\sin^2\alpha\sin^2\theta\right)\nonumber\\
&-&\frac{1}{2}\left( e_{11}+\beta_3-e_{33}\right) \sin\alpha\sin^2\theta\,qE-\frac{1}{4} \varepsilon_0\varepsilon_a\,E^2\sin^2\theta\nonumber\\
\end{eqnarray}}
where $\widetilde{K}_{11}=K_{11}+2\nu$, with $\nu=\nu_1+\nu_2+\nu_3+\nu_5$.
Note that in $<f_{tb}>$ expression appears a new contribution to the mean flexoelectric energy term, namely the $\beta_3$-term  arising from the twist-bend cone distortion by the action of the electric field, since for $\alpha\ne 0$ the director acquires a z-dependence  $n_z=n_z(z)$. The dielectric torque stabilizes the helix axis for $\varepsilon_a < 0$. For simplicity, in the subsequent analysis, we consider a system with $\varepsilon_a=0$, that is, the helix axis orientation is fixed. For the needs of numerical analysis we use typical values for the elastic constants $K_{11}\simeq 8\,\mathrm{pN}$, $K_{22}\simeq 3\,\mathrm{pN}$, $K_{33}\simeq 0.5\,\mathrm{pN}$ close to what was measured for CB7CB \cite{olegK}. For $E=0$, the pitch is taken $p_0 \simeq 10\,\mathrm{nm}$ \cite{zhu} and the tilt angle of the undisturbed helix $\theta_0\simeq 20^{\circ}$.

\section{Results}

Minimization of Eqn (\ref{fmean}) in respect to the wavevector $q$ and to the order parameter $x=\sin^2\theta$ results to the following expressions for the equilibrium wavevector $q_{\alpha}$ of the modulation
\begin{equation}
\label{q}
q_{\alpha}=\pm\sqrt{\frac{\eta(2\cos^2\alpha-\sin^2\alpha)}{\Pi_q}}
\end{equation}
where $\pm$ means that, unlike the case of N*, formation of $N_{TB}$ does not require molecular chirality thus  domains of left and right chirality coexist.
The order parameter value  of  the  electric field distorted $N_{TB}$ phase at equilibrium  is
\begin{eqnarray} \label{x}
x_{\alpha} &=&\frac{ 4 q_{\alpha} \kappa_2\cos\alpha+2q_{\alpha} e_{ef} E\sin\alpha-2 \eta(2\cos^2\alpha-\sin^2\alpha)}{q_{\alpha}^2\Pi_x}\nonumber\\
\end{eqnarray}
where
\begin{eqnarray*}
\Pi_x &=& 2(K_{22}-K_{33})(2\cos^2\alpha-\sin^2\alpha)\\
&+&\mu_1\sin^2\alpha(\sin^2\alpha-4\cos^2\alpha)	\\
\Pi_q &=& (\widetilde{K}_{11}+K_{22})\sin^2\alpha + 2 K_{33}\cos^2\alpha+4\nu_4\\
&+& 2\mu_1\cos^2\alpha\sin^2\alpha\\
e_{ef} &=& e_{1}+\beta_3-e_{3}
\end{eqnarray*}
These expressions for $q_{\alpha}$ and $x_{\alpha}$ in the presence of flexoelectricity, generalize Eqn(13-14) of Ref\cite{pre}. From  Eqn(\ref{x}) we infer that the effect of the flexoelectric coupling  on the heliconical angle variation for domains with opposite chirality is the same. For the subsequent analysis we chose for the chirality the "+" sign without loss of generality.

\begin{figure}[h]
	\centering	
\includegraphics[width=8cm]{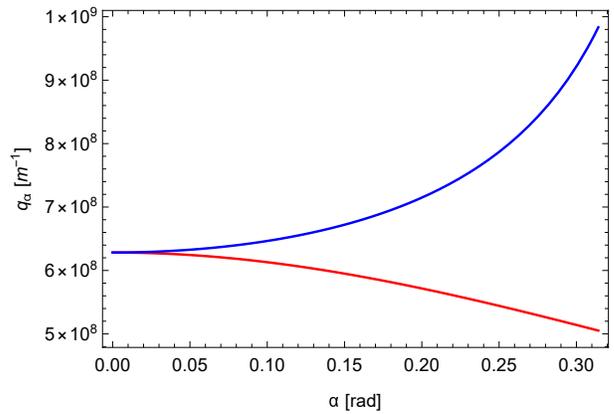}
	\caption[]{Equilibrium wavevector $q_{\alpha}$ vs optical axis rotation angle $\alpha$ caused by the flexoelectric coupling. Upper branch (blue line) where $p_{\alpha}\sim 1/E$, for $K_{33}+2\nu_4 < 0$ and  $\nu_4=-K_{22}/4$. Lower branch (red line) where $p_{\alpha}\sim E$, for $K_{33}+2\nu_4 > 0$ and  $\nu_4=K_{22}/4$.}
	\label{Figure_1}
\end{figure}
In Figure 1, we show $q_{\alpha}=q(\alpha)$ behavior which exhibits two regimes depending on the sign of the effective bend elastic constant $K_{33}+2\nu_4$, as it can be analytically shown by expanding $q_{\alpha}$ up to $\alpha^2$ terms. This expansion yields
%%%%%%%%%%%%%%%%%%%%%%%%%%%%%%%%%%%%%%%%%%%%%%%%
\begin{eqnarray}
q_{\alpha} = q_0\left(1-\frac{ K_{\text{ef}}+K_{22}+K_{33}+
	2 \mu _1+6 \nu _4}{4 \left(K_{33}+2 \nu _4\right)}\,\alpha ^2\right)+\,\mathcal{O}(3)\nonumber\\
\end{eqnarray}
where $q_0$ is the wavevector of the undistorted structure (in the absence of electric field) and $K_{ef}= \widetilde{K}_{11}+(1-3 x)K_{22} +(3 x-2)K_{33}$ is an effective elastic constant. The last expression indicates that the wavevector may be reduced or increased with the rotation angle $\alpha$ or equivalently with the electric field $E$, depending mostly on the sign of the $K_{33}+2 \nu _4$ term.

 For $K_{33}+2\nu_4>0$, $q$ decreases with $\alpha$ (lower red curve in Fig.1) while in the opposite case $q$ increases (upper blue curve) with $\alpha$.

In order to obtain the dependance of the rotation angle $\alpha$ on $E$, we further minimize the energy in respect to $\alpha$ and obtain the following expression
\begin{eqnarray}\label{E(a)}
	e_{ef} E  x_{\alpha} &=& (2-3 x_{\alpha})\sin\alpha\, \frac{\eta}{q_{\alpha}}+2 x_{\alpha} \kappa_2 \tan\alpha \nonumber\\
	&+&	q_{\alpha}x_{\alpha}\sin\alpha\left\{ K_{ef} +\frac{1}{2}\mu _1 \left[ x_{\alpha}+(4-5x_{\alpha}) \cos (2\alpha)\right]\right\}\nonumber\\
\end{eqnarray}
As expected, when $E$ changes sign then $\alpha$ changes sign too.
The system of Eqs (\ref{q}, \ref{x}, \ref{E(a)}) can be resolved both analytically  and numerically to obtain $x(\alpha)$ and $E(\alpha)$. Nevertheless, the analytical solution is cumbersome and it is not presented here. Therefore in order to obtain simple analytic expressions, we introduce the approximation of small rotation angle.
For $\alpha<<1$ and omitting terms in $\mathcal{O}(\alpha^2)$ the above Eqn(\ref{E(a)}) simplifies to	
\begin{equation}	\label{small_a}
\alpha = \frac{e_{ef} E q_{\alpha} x_{\alpha}}{q_{\alpha}^2 x_{\alpha} \widetilde{K}_{ef} +  2 q_{\alpha} x_{\alpha} \kappa_2+ (2-3 x_{\alpha})\eta}
\end{equation}
where $\widetilde{K}_{ef}={K}_{ef}+2\mu_1(1-x)$
 is an elastic constant. Note that all bulk elastic constants are present in the expression of $\widetilde{K}_{{ef}}$.  The linear approximation for $E(\alpha)$  works well for the experimental situation ($\alpha\sim 1^{\circ}$, \cite{claire}) and  it starts to deviate from the exact solution for $\alpha \gtrsim 10^{\circ}$.

Expanding Eqn(\ref{x}) in powers of $\alpha$ with the use of Eqn(\ref{small_a}) we find that the order parameter variation, $\delta x(\alpha)$, has no linear term in $\alpha$. The first non zero term is the quadratic one  $\delta x\sim\alpha^2+\dots$. Thus, the tilt angle variation does not depend on the sign of chirality.  In Figure 2 we show the $\theta (\alpha)$ dependence. The approximation of constant $\theta=\theta_0$ seems fair for small enough rotations of the optical axis. For the experimentally observed values of $\alpha$ \cite{claire}, corrections are of the order of a few per cent. For $\alpha >>1^{\circ}$ the heliconical angle variation becomes large.

\begin{figure}[h]
	\centering	\includegraphics[width=8cm]{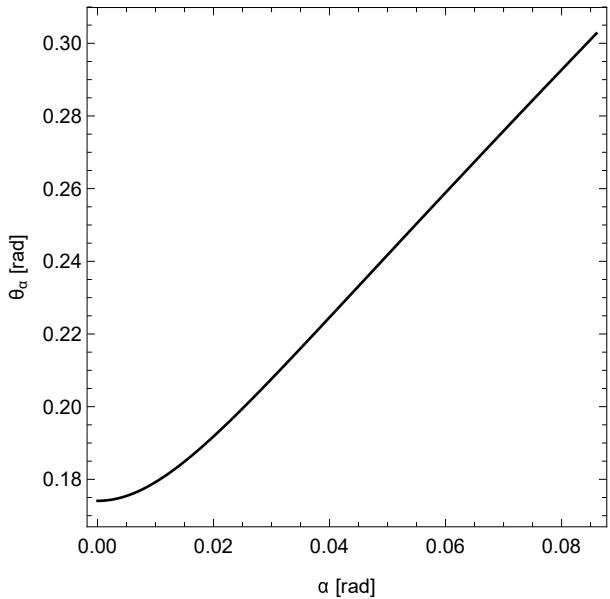}
	\caption[]{Equilibrium conical angle $\theta_{\alpha}$ vs optical axis rotation angle $\alpha$ induced by a transverse electric field numerically calculated.}
	\label{Figure_3}
\end{figure}

In conclusion, for $\alpha\lesssim 1^{\circ}$ one can assume that $q_{\alpha}=q_0$ and $x_{\alpha}=x_0$. Then Eqn(\ref{small_a}) further simplifies and can be cast to the familiar expression for cholesteric liquid crystals
\begin{equation}	\label{a-s}
\alpha =  \frac{e_{fl} E}{q_0\,K(\theta_0)}
\end{equation} where $K(\theta)=\widetilde{K}_{ef}  + 2(K_{22}+2\nu_4)+ (K_{33}+2\nu_4) (-3+2/x)$ is a global elastic constant that depends from the set of elastic constants entering Eqn(\ref{fmean}) and on the heliconical tilt angle. The latter dependence results to a rotation angle that goes to zero as the heliconical tilt angle vanishes.

Finally, we complete the present analysis with the calculation of the characteristic relaxation time $\tau_{off}$ for the linear electrooptic effect.  $\tau_{off}$ can be calculated by considering the equilibrium equation of the elastic and viscous torques, that results to the following expression

\begin{equation}	\label{t-s}
\tau_{off}=\frac{2\gamma_1}{K(\theta_0)\sin^2\theta_0 q_0^2}
\end{equation} where $\gamma_1$ is the nematic rotational viscosity coefficient.
For  $\theta_0=\pi/2$, Eqn(\ref{a-s},\ref{t-s}) reduce to the corresponding equations of the cholesteric phase \cite{Corbett}.
%%%%%%%%%%%%%%%%%%%%%%%%%%%%%%%%%%%%%%%%%%%%%%%%
%%%%%%%%%%%%%%%%%%%%%%%%%%%%%%%%%%%%%%%%%%%%%%%%
%%%%%%%%%%%%%%%%%%%%%%%%%%%%%%%%%%%%%%%%%%%%%%%%
\section{Discussion \& Conclusions}
%%%%%%%%%%%%%%%%%%%%%%%%%%%%%%%%%%%%%%%%%%%%%%%%

The above analysis results in some interesting conclusions. First, the sign of the effective bent elastic constant $K_{33}+2\nu_4$ results in qualitatively opposite behavior of the pitch variation caused by the flexoelectric effect.  For $K_{33}+2\nu_4<0$, the pitch decreases with increasing electric field and one retrieves the result for the flexoelectric effect in cholesteric liquid crystals \cite{Corbett} where flexoelectric coupling opposes the effect of dielectric coupling. The same dependence, $p\sim 1/E$ was shown in the case of oblique helicoidal structures induced by an electric field in a cholesteric phase \cite{oleg}.
In the opposite case for $K_{33}+2\nu_4>0$ where the pitch increases with the electric field, the flexoelectric effect starts to unwind the twist-bend helix as in the case of the cholesteric helix unwinding by dielectric coupling \cite{durand,prost}. Therefore one could, in principle, unwound the helix via a transition towards a distorted nematic phase under field. This latter nematic should be a distorted splay-bend nematic since the twist is excluded by the field. Such an nematic phase was predicted both as a ground state and as a field stabilized phase in \cite{dozov,splaybend,longa,moliq}.  Apparently, a measurement of the pitch variation sign caused by flexoelectricity will give information about the sign of the $K_{33}+2\nu_4$ elastic constant.

Recently, Balachandran et al \cite{balachandran} experimentally found for the bimesogenic liquid crystal $CB11CB$ that its flexoelectric polarization in the $N_{TB}$-phase, is at least two times larger than in its own classical $N$-phase. Their measurements gave an effective flexoelectric coefficient $\left|e_{ef} \right|\sim 35\,\mathrm{pC/m}$ deep in the $N$-phase. A similar value, $\sim 30\,\mathrm{pC/m}$, was found for CB7CB in \cite{Zhou}. Typically, $\left|e_{ef} \right|\sim 1-10\,\mathrm{pC/m}$ for rod-like monomesogens \cite{buka,BlinovPRE00,5cb} while it can be larger for bent-core mesogenic molecules \cite{kaur} and  for dimeric rod-like mesogens \cite{balachandran}. In our knowledge, up to now, no model predicts a further increase of the effective flexoelectric coefficient once a system undergoes the uniaxial nematic to the twist-bend nematic phase transition. According to our model, the measured increase of the effective flexoelectric polarization could be justified by the activation of the $\beta_3$-mode at the transition towards the $N_{TB}$-phase.  Concerning the nature of the $\beta_3$-term, from Eqn(\ref{totalfl}) is inferred that it is inherently related to the formation of the helix and is activated from gradients, parallel to helix-axis, of the heliconical tilt angle. Therefore the $\beta_3$-term is not simply related to the molecular structure of the mesogens but also to the deformation of their collective helix geometry.

Finally, we investigate the experimental data for the flexoelectric effect in CB7CB \cite{claire} employing the present model. Using the reported values for the electric field $E=25\,\mathrm{V/\mu m}$, rotation angle $\alpha\simeq 1.1^{\circ}$, $\tau \lesssim 1\mathrm{\mu s}$ \cite{claire} and  $\left|e_{ef} \right|\simeq 35\,\mathrm{pC/m}$ \cite{balachandran, Zhou}, we find a pitch of $\simeq 10 \,\mathrm{nm}$ close to the transition temperature for a tilt angle of $\theta\simeq 12^{\circ}$ in good agreement with the experimentally deduced values \cite{zhu}. Note that if the dependence of the elastic constants on the heliconical tilt angle is not taken into account, then the predicted value for the pitch is about 2nm.

In conclusion, we demonstrated that in the twist-bend nematic phase a new contribution to the flexoelectric response arises caused by the gradient of the heliconical angle induced by a transverse to the helical axis electric field. This new contribution justifies the observed increase of the flexoelectric polarization measured in \cite{balachandran}. We generalized the expression of the optical axis rotation angle for the cholesteric phase in order to take into account the dependence of the elastic constants on the heliconical angle. The proposed model gives the correct value for the heliconical pitch in the $N_{TB}$-phase. Finally, an electric field induced splay-bend nematic phase was predicted.

%\newpage

%%%%%%%%%%%%%%%%%%%%%%%%%%%%%%%%%%%%%%%%%%%%%%%%%%%%%%%%%%%%%%%%%%%%%%%%%%%%%%%%%%%%%%%%%%%%%%%%

\end{document}